\documentclass[aip,graphicx]{revtex4-1}

\usepackage{graphicx}
\usepackage{xcolor}

\begin{document}

\title{\textcolor{black}{Scalable interdigitated photoconductive emitters for the electrical modulation of terahertz beams with arbitrary linear polarization}}

\author{C.~D.~W.~Mosley}
\affiliation{Department of Physics, University of Warwick, Gibbet Hill Road, Coventry CV4 7AL, UK}

\author{M. Staniforth}
\affiliation{Department of Physics, University of Warwick, Gibbet Hill Road, Coventry CV4 7AL, UK}
\affiliation{Department of Chemistry, University of Warwick, Gibbet Hill Road, Coventry CV4 7AL, UK}

\author{\textcolor{black}{A.~I.~Hernandez Serrano}}
\affiliation{Department of Physics, University of Warwick, Gibbet Hill Road, Coventry CV4 7AL, UK}

\author{\textcolor{black}{E.~Pickwell-MacPherson}}
\affiliation{Department of Physics, University of Warwick, Gibbet Hill Road, Coventry CV4 7AL, UK}

\author{J.~Lloyd-Hughes}
\email[]{j.lloyd-hughes@warwick.ac.uk}
\affiliation{Department of Physics, University of Warwick, Gibbet Hill Road, Coventry CV4 7AL, UK}

\date{20 December 2018}

\begin{abstract}
A \textcolor{black}{multi-element} interdigitated photoconductive emitter for broadband THz polarization rotation is proposed and experimentally verified. The device consists of separate pixels for the emission of horizontally and vertically polarized THz radiation. The broadband $(0.3-5.0$\,THz$)$ nature of the device is demonstrated, and the polarization angle of the generated far-field THz radiation is shown to be readily controlled by varying the relative bias voltage applied to the horizontally and vertically emitting pixels. The device is scalable in design, and with its simple method of polarization rotation it allows the modulation of the generated THz polarization at rates significantly faster than those acheivable in ellipsometry systems based on mechanically rotating components.
\end{abstract}


\maketitle

Photoconductive emitters are a crucial component for terahertz (THz) radiation sources and detectors, being the most commonly used in commercially available and custom-built spectroscopy and imaging systems based on oscillator and fiber lasers.\cite{Castro-Camus2016,Burford2017} A recent development in THz generation from photoconductive emitters is the creation of complex polarization states,\cite{Winnerl2009,Kan2013,Waselikowski2013,Cliffe2014} and the ability to control the polarization state of the emitted radiation.\cite{Castro-Camus2005b,Hirota06-1552,Mosley2017} These advances in THz polarization control may permit otherwise inaccessible excitations, and the anisotropic properties of materials, to be probed in the THz region.

Some photoconductive emitter designs, such as dipole antenna arrays\cite{Froberg1992,Berry2014} and interdigitated photoconductive emitters,\cite{Dreyhaupt2005,Hattori2006} create a number of individual dipoles that interact in the far-field to improve the generation efficiency of linearly polarized THz radiation. For complex polarization states, often the generation schemes utilize emitter geometries that create a pattern of individual dipoles with different orientations, which then interact in the far-field to create the desired polarization state. Radially- and azimuthally-polarized THz beams have been generated using interdigitated photoconductive emitters with suitable electrode geometries,\cite{Winnerl2009,Kan2013} and radially polarized beams have been produced by circular large-gap emitters.\cite{Waselikowski2013,Cliffe2014} Control of the THz polarization state has been demonstrated by rotating a wide-gap photoconductive emitter to a few fixed angles,\cite{Castro-Camus2005b} and modulation between two orthogonal polarization states has been achieved with a four-contact large area emitter.\cite{Hirota06-1552} A method of rotating a linear polarization state to arbitrary angles has been demonstrated by the mechanical rotation of an interdigitated photoconductive emitter,\cite{Mosley2017} which gave an accuracy and precision in the angle of the emitted radiation comparable to leading THz ellipsometry techniques.\cite{Neshat2013} 

Here we report the design, fabrication and validation of interdigitated photoconductive emitters consisting of separate pixels for the generation of horizontally and vertically polarized THz radiation. This photoconductive emitter design has the benefit of allowing the direct control of the polarization angle of the emitted THz beam, without requiring any additional or mechanically moving components, via the applied bias voltage on the pixels. 

\begin{figure}[t]
\includegraphics[width=0.5\textwidth]{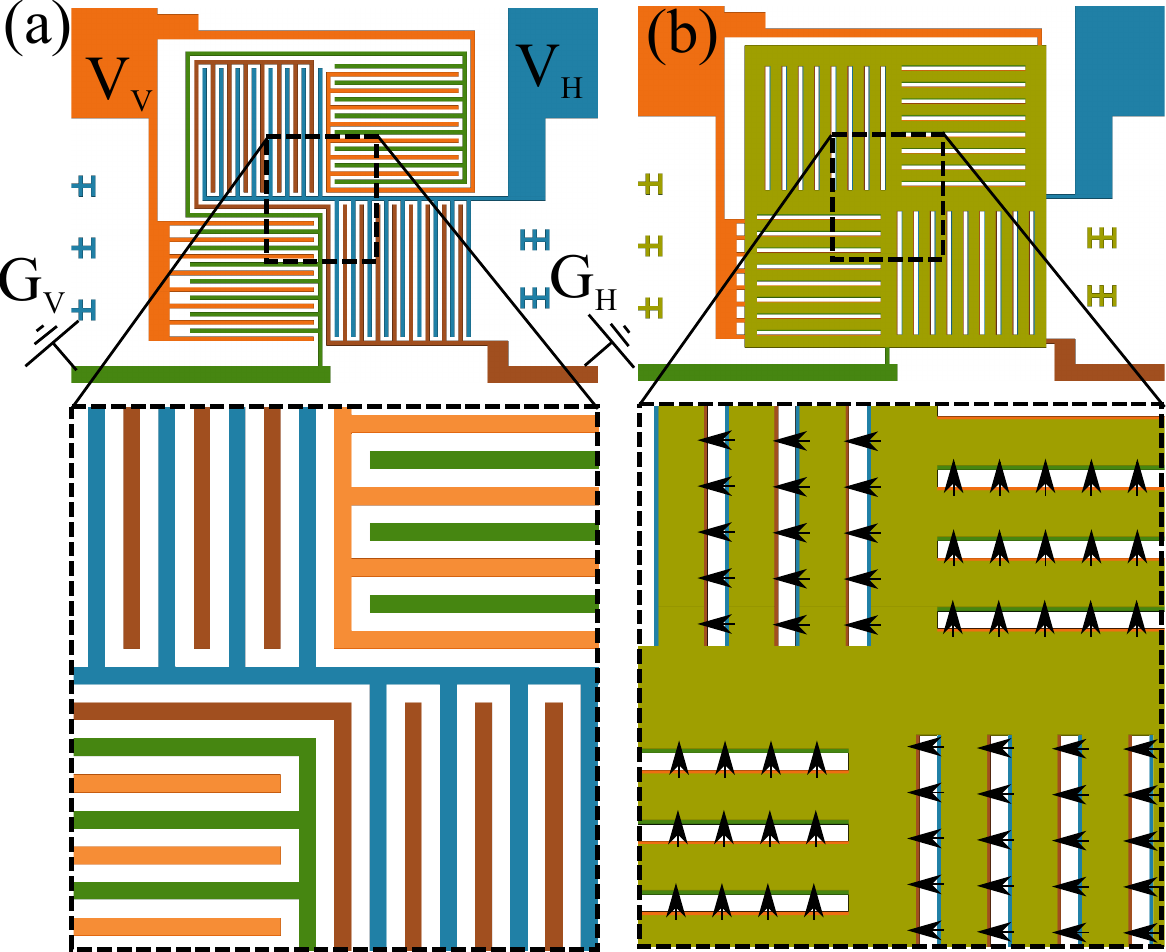}%
\caption{\label{Figure1} Schematic diagram of the multi-pixel interdigitated photoconductive emitter, showing \textbf{(a)} the interdigitated electrodes and \textbf{(b)} the completed device with the masking layer. $V_{\rm{H}}$ ($G_{\rm{H}}$) and $V_{\rm{V}}$ ($G_{\rm{V}}$) signify the biased (grounded) contacts for horizontally and vertically polarized emission, respectively. The direction of the bias field, and therefore polarization state of the emitted THz pulse, is demonstrated by the arrows in \textbf{(b)} for emission at at a target polarization angle of $135^{\circ}$.}%
\end{figure}

The geometry of the multi-pixel interdigitated photoconductive emitter is shown schematically in Figs.\,\ref{Figure1}(a) and \ref{Figure1}(b). The layout of the device is a $2\times2$ grid of pixels, with the electrodes in each pixel arranged orthogonally to those in the adjacent pixels. In a photoconductive emitter, the direction of the bias field applied between the electrodes defines the direction of acceleration of the photoexcited charge carriers, and therefore the polarization state of the generated THz radiation, demonstrated by the arrows in Fig.\,\ref{Figure1}(b). The pixels producing horizontally polarized radiation share a common set of bias and ground electrodes, indicated in Fig.\,\ref{Figure1}(a) by $V_{\rm{H}}$ and $G_{\rm{H}}$ respectively, as do the vertically emitting pixels ($V_{\rm{V}}$ and $G_{\rm{V}}$ in Fig.\,\ref{Figure1}(a)). Photoexciting the entire active area of the device and varying the relative strengths of $V_{\rm{H}}$ and $V_{\rm{V}}$ will vary the relative strengths of the horizontally and vertically polarized components of the generated THz radiation; hence the orientation of the polarization state of the far-field THz pulse generated by the device can be defined by simply varying the relative bias voltage on the horizontal and vertical contacts. 

\begin{figure}[tb]
  \centering
  \includegraphics[width=0.9\textwidth]{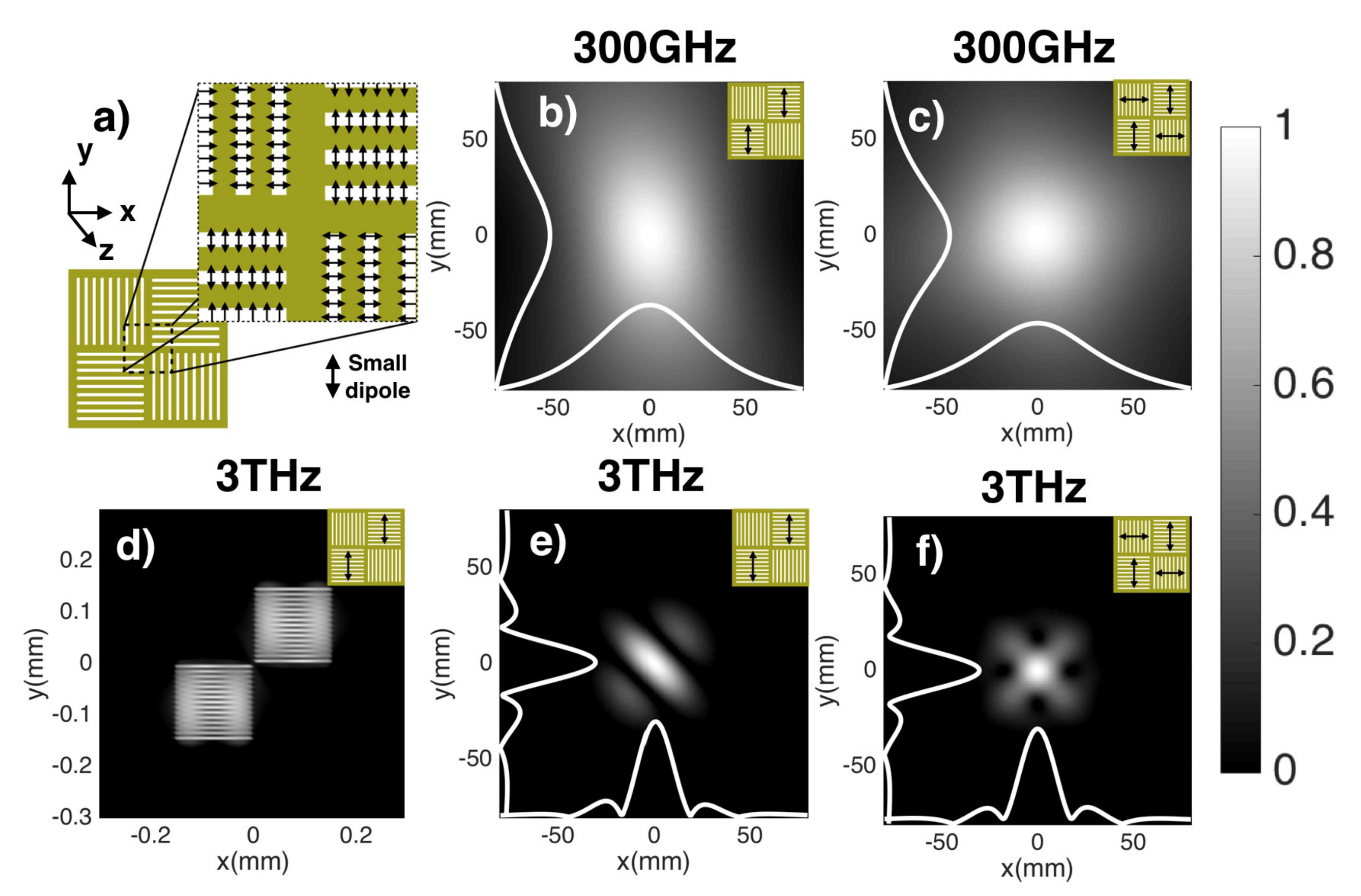}
\caption{\textcolor{black}{Calculated radiation patterns in the $x-y$ plane for the multi-pixel interdigitated emitter. (a) Diagram of the numerical simulation. (b) Far-field ($z=50$\,mm) THz electric field amplitude $E$ at 300\,GHz with only the vertically-emitting pixels active. (c) As (b), but with both vertical and horizontal pixels on. (d) Near-field $E$ at 3\,THz with only the vertical pixels on ($z=5$\,$\mu$m). (e) Far-field ($z=50$\,mm) $E$ at 3\,THz with vertically-emitting pixels active. (e) As (d), with both vertical and horizontal pixels on.}}
\label{Figure-theory}
\end{figure}
 
\textcolor{black}{To explore the expected far-field performance of the device, we calculated the radiation pattern of an array of dipoles arranged with the same spatial distribution as the active semiconductor area of the device. Point dipoles were placed at 5\,$\mu$m intervals along each active finger, as indicated in Fig.\ \ref{Figure-theory}(a) by the small arrows. For an individual point dipole, the electric field vector can be calculated analytically at an arbitrary location.\cite{Stutzman2013} The total THz electric field amplitude $E(x,y,z)=\sum_i\sqrt{E^2_{x,i}+E^2_{y,i}}$ was then calculated at each position $x, y$ (in an image plane a distance $z$ from the emitter) by summing the contribution from all dipoles in the array, each with index $i$. Each dipole was assumed to have the same strength, corresponding to uniform excitation of the emitter. }

\textcolor{black}{Here we present calculations for two frequencies representative of the broadband spectrum produced by a typical photoconductive emitter, which peaks around 1\,THz.} In Fig.\ \ref{Figure-theory}(b) $E(x,y)$ is shown at 300\,GHz at a distance $z=50$\,mm away from the device, for the case of vertical emission only (as indicated by the inset). This $z$ position corresponds to the effective focal length of the first (collimating) off-axis parabolic mirror in the experiment. The radiation pattern in the far-field has gaussian cross-sections in $x$ and $y$ (white lines), is close to circular, and becomes more circular when both horizontal and vertical pixels are excited [panel (c)]. In contrast, in panel (d) the near-field $E(x,y,z=5$\,$\mu$m$)$ at 3\,THz and for vertical emission is shown, where each photoexcited dipole line can be seen. As the radiation evolves into the far-field, the individual lines are no longer visible, and the radiation pattern becomes more beam-like for vertical [panel (e)] and combined [panel (f)] emission, with a beam width below 40\,mm at $z=50$\,mm. However as the divergence of a higher frequency beam is lower, the beam profiles at 3\,THz are more non-uniform than at 300\,GHz, and exhibit some areas of destructive interference. Further calculations may suggest design improvements to mitigate against these effects, for instance using a smaller pixel size.

The multi-pixel interdigitated photoconductive emitter was realised via UV photolithography and metal evaporation on a 500\,$\mu$m-thick semi-insulating GaAs substrate. The bottom layer of the device, shown in Fig.\,\ref{Figure1}(a), consists of 300\,nm-thick, 5\,$\mu$m-width Au electrodes, with a 5\,$\mu$m gap between electrodes. To avoid destructive interference of the generated THz radiation in the far-field, a 300\,nm-thick Au masking layer was deposited on top of the metal electrodes, shown in Fig.\,\ref{Figure1}(b), to block the photoexcitation beam in half of the gaps. To avoid short circuiting, the electrodes and masking layer were vertically separated by a 110\,nm-thick insulating layer of Al$_2$O$_3$, with such a thickness chosen in order to maximize the transmission of the 800\,nm photoexcitation beam. The active area of each pixel is defined by the masking layer, and is $150$\,$\mu$m$\times150$\,$\mu$m. \textcolor{black}{The approach is scalable to larger or smaller pixel sizes, or to a higher number of pixels, simply by using alternative UV photolithography masks.}

The performance of the fabricated device was investigated by terahertz time-domain spectroscopy. The emitter was excited at 800\,nm with a 350\,mW beam from an 80\,fs pulse duration, 80\,MHz repetition rate Ti:Sapphire laser oscillator. The bias voltages applied to the horizontal and vertical contacts were varied independently from zero to a maximum voltage of $\pm10$\,V, with the voltage source modulated at a frequency of 50\,kHz. The spectrometer consisted of 4 off-axis parabolic mirrors, with the mirror collecting the generated THz radiation having a focal length $f=50.8$\,mm, the following two mirrors having focal lengths $f=76.2$\,mm, and the mirror focussing onto the detector having a focal length $f=101.6$\,mm. Each successive mirror in the beam path was oriented such as to create a ``z-shape'' spectrometer. The generated THz pulses were detected via polarization-resolved electro-optic sampling \cite{VanderValk2004,vanderValk05-2839} in a 200\,$\mu$m-thick, $\langle111\rangle$-surface normal GaP crystal. This variation on standard electro-optic sampling permits two orthogonal components of the THz pulse to be detected, allowing the full THz polarization state to be used for THz spectroscopy\cite{Lloyd-Hughes2014,Failla2016,Mosley2017} and imaging.\cite{vanderValk05-2839} \textcolor{black}{The peak THz electric field strength was 3.4\,kVm$^{-1}$ from electro-optic sampling, in agreement with the average power of 1\,$\mu$W measured by a calibrated pyroelectric. Comparable emission strengths were obtained from reference devices with only one large pixel.} Measurements were performed under a dry nitrogen purge, in order to avoid atmospheric absorption of the THz radiation. 

The polarization state of the generated THz pulses was parameterised by their ellipticity, $\chi$, and orientation angle, $\psi$. An ellipticity of $0^{\circ}$ corresponds to a linear polarization state, whilst $\chi=\pm45^{\circ}$ corresponds to right- and left-handed circularly polarized states, respectively. The orientation angle is the relative angle between the polarization state and the $x$-axis of the lab reference frame, which is defined by the in-plane axes of the detection crystal. $\chi$ and $\psi$ were obtained from the experimental data by converting the complex THz spectra $\widetilde{E}_x(\omega)$ and $\widetilde{E}_y(\omega)$ into a circular basis using
\begin{equation}
 \widetilde{E}_{\pm}=|\widetilde{E}_{\pm}|e^{i\phi_{\pm}}=\frac{\widetilde{E}_x{\pm}i\widetilde{E}_y}{\sqrt{2}},
\label{Eq1}
\end{equation}

\noindent
and by then using
\begin{equation}
 \tan\,(\chi)=\frac{|\widetilde{E}_-|-|\widetilde{E}_+|}{|\widetilde{E}_+|+|\widetilde{E}_-|},
\label{Eq2}
\end{equation}

\noindent
and
\begin{equation}
 \psi=\frac{\phi_+-\phi_-}{2}.
\label{Eq3}
\end{equation}

\begin{figure}[t] 
\includegraphics[width=0.8\textwidth]{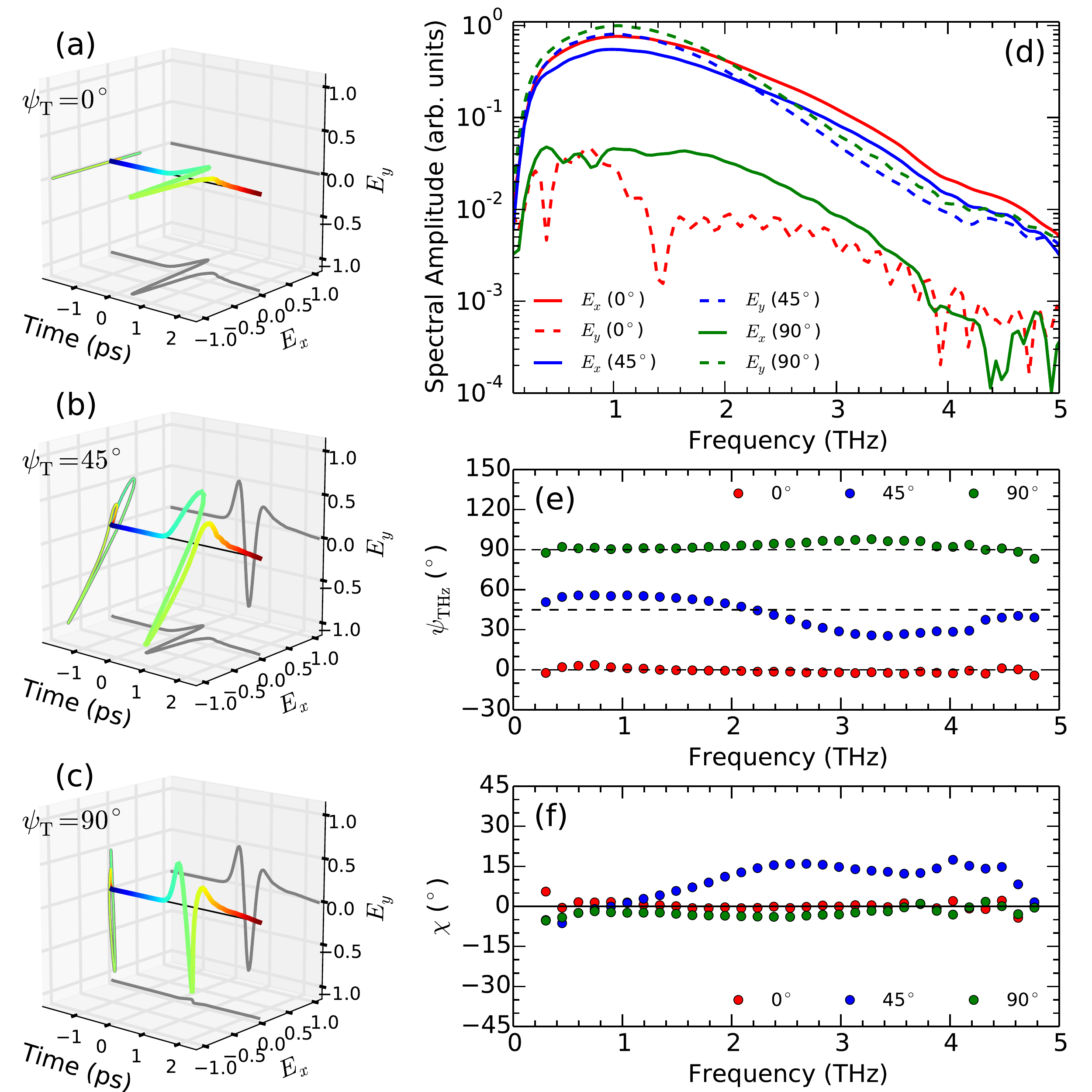}%
\caption{\label{Figure2} Polarization-resolved time-domain traces for THz pulses emitted from the multi-pixel interdigitated photoconductive emitter at target polarization angles of \textbf{(a)} $\psi_{\rm{T}}=0^{\circ}$, \textbf{(b)} $\psi_{\rm{T}}=45^{\circ}$ and \textbf{(c)} $\psi_{\rm{T}}=90^{\circ}$. \textbf{(d)} The corresponding Fourier transform spectra of each component to the time-domain traces in \textbf{(a)}-\textbf{(c)}. \textbf{(e)} and \textbf{(f)} show the orientation angle and ellipticity, respectively, at each target angle.}%
\end{figure}

To initially verify that the pixel emitter works as intended, three cases of emitter bias were tested: biasing the horizontally emitting pixels only, to produce a THz pulse with a target polarization angle $\psi_{\rm{T}}=0^{\circ}$; applying the same bias voltage to both sets of pixels, for $\psi_{\rm{T}}=45^{\circ}$; and biasing the vertically emitting pixels only, for $\psi_{\rm{T}}=90^{\circ}$. Polarisation-resolved time-domain traces are shown in Fig.\,\ref{Figure2}(a)-(c), respectively. In each case, the emitter produced linearly polarized pulses of THz radiation, and the polarization state of the pulses was rotated as intended by changing the bias voltage applied to the device.

The corresponding Fourier transform spectra of the time-domain traces are shown in Fig.\,\ref{Figure2}(d). In each of the three cases, the generated THz pulses demonstrate broadband frequency components from $0.3-5.0$\,THz. To quantify the polarization state of the generated THz pulses, their frequency-dependent ellipticity and orientation angle were extracted from the spectra using equations \ref{Eq1}-\ref{Eq3}, and are shown in Figs.\,\ref{Figure2}(e) and \ref{Figure2}(f) respectively. In the $0^{\circ}$ and $90^{\circ}$ cases, there is only a small ellipticity present and the orientation angle of the pulses remain close to the target values (represented by the dashed lines), demonstrating that the generated pulses are linearly polarized over the entire broadband frequency range. In the $45^{\circ}$ case, there is more variation in both the ellipticity and the orientation angle, particularly at higher frequencies, however the polarization state has clearly been rotated in between the $0^{\circ}$ and $90^{\circ}$ cases.

\begin{figure}[t]
\includegraphics[width=0.8\textwidth]{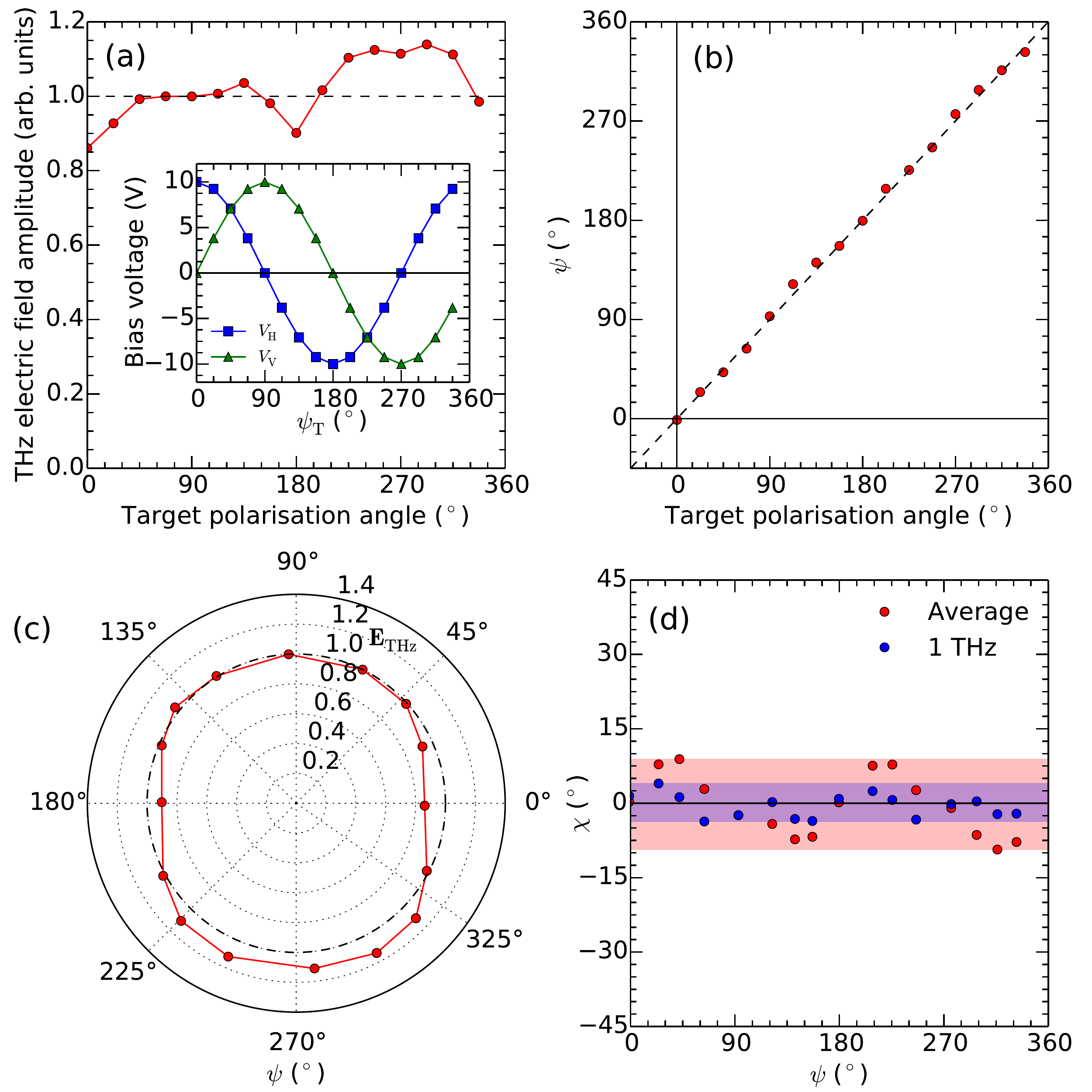}  
\caption{\label{Figure3} \textbf{(a)} Maximum THz electric field amplitude, normalised to the mean value, as the polarization angle is varied over a $360^{\circ}$ range by changing the relative bias voltages between horizontal and vertical pixels. The inset shows the bias voltage applied for each target polarization angle. \textbf{(b)} Comparison of the experimentally measured orientation angle to the target angle at each step. The dashed line represents an exact match between the two. \textbf{(c)} Polar representation of the orientation angle and amplitude of the THz pulses at each step. \textbf{(d)} Ellipticities of the generated THz radiation at each orientation angle, at 1\,THz (blue) and averaged from 0.3-5.0\,THz (red). The shaded areas represent the variation over the $360^{\circ}$ rotation.}
\end{figure}

Having proved the principle of rotating the polarization state of the emitted THz pulses by varying the relative bias voltage applied to each set of contacts, the characteristics of the emitted THz radiation were investigated as the orientation angle of the polarization state was varied over a $360^{\circ}$ range. In order to rotate the polarization state, the bias voltages applied to the horizontally and vertically emitting pixels were varied according to $V_\mathrm{H}=V_\mathrm{max}\cos(\psi_{\mathrm{T}})$ and $V_\mathrm{V}=V_\mathrm{max}\sin(\psi_{\mathrm{T}})$, 
%
%
%
where $V_{\rm{max}}=10$\,V and $\psi_{\rm{T}}$ is the target orientation angle, and is shown in the inset of Fig.\,\ref{Figure3}(a). The target polarization angle was varied in $22.5^{\circ}$ steps and polarization-resolved time-domain traces taken at each step.

The maximum amplitude, $E_{\rm{max}}=(E_{x}^{2}+E_{y}^{2})^{1/2}$, at the time-domain peak of the THz pulse at each target angle is reported in Fig.\,\ref{Figure3}(a), normalised to the mean value. The variation in amplitude is less than $\pm15\%$ over the $360^{\circ}$ range, demonstrating that the device maintains a fairly consistent THz emission strength at all angles. Figure \ref{Figure3}(b) shows the orientation angle of the emitted THz pulses, averaged over their $0.3-5.0$\,THz bandwidth, versus the target orientation angle. The measured orientation angle remains close to the target angle for all points, with the ideal case represented by the dashed line, demonstrating that the orientation angle of the emitted radiation can be reliably varied by changing the relative bias voltage on the horizontally and vertically emitting pixels.

The combined data from Figs.\,\ref{Figure3}(a) and \ref{Figure3}(b) are presented in Fig.\,\ref{Figure3}(c). The THz amplitude is observed to be larger for vertical emission than for horizontal emission, and also to increase when the bias voltage applied to the contacts is negative rather than positive. The increase in THz emission strength when negatively biasing the contacts occurs due to the well-known effect in photoconductive antennas whereby exciting the device closer to the anode increases the THz emission strength.\cite{Ralph1991,Gregory2005,Castro-Camus2005} In the case of negative bias on the multi-pixel emitter, excitation over the entire device acts as being close to the anode, producing more efficient THz emission than with a positive bias. The difference in response between the horizontal and vertical emission may occur due to the polarization of the pump pulse relative to the geometry of the pixels; it has been previously shown that orienting the polarization state of the pump pulse parallel to the wires of an interdigitated photoconductive emitter increases the THz emission strength.\cite{Mosley2017} In the setup used in this work, the polarization of the pump was parallel to the wires in the vertically emitting pixels, enhancing their emission strength relative to the horizontally emitting pixels. The amplitude of the THz pulses emitted by the device presented here may therefore be made more uniform during rotation of the polarization state by: using a half-wave plate to rotate the polarization of the pump pulses to $45^{\circ}$, to ensure a similar response from both sets of pixels; and by using only negative bias voltages and swapping the biased and grounded contacts to change the polarity of the generated THz pulses.

The ellipticities of the generated THz radiation at each orientation angle are reported in Fig.\,\ref{Figure3}(d). The red points represent the ellipticity averaged over the $0.3-5.0$\,THz bandwidth of the pulses, whilst the blue points  show the ellipticity at 1\,THz. At 1\,THz, the ellipticity remains small at all angles, varying less than $4^{\circ}$ over the full $360^{\circ}$ rotation. When the higher frequency components are taken into account, the variation in ellipticity becomes larger, particularly when the polarization state is between a purely horizontal or vertical state. One reason for the increase in ellipticity at these angles may be due to the different response of the pixels due to their geometry relative to the polarization of the pump pulse. The grating created by the interdigitated wires of the emitter has a different reflectivity for an s- or p-polarized pump beam, and hence different pump beam powers are coupled to the photoconductor, giving a different carrier density in each set of pixels. Such an effect may be observed in Fig.\,\ref{Figure2}(d), where the horizontal and vertical components of the electric field in the $\psi_{\rm{T}}=45^{\circ}$ case have slightly different spectral shapes, as a result of small differences in the time-domain waveforms for the two components. Another reason for the larger variation at higher frequencies may be due to the higher frequencies having a smaller beam divergence, and as such creating a slightly asymmetrical beam profile in the far-field if the high frequency components generated from each pixel do not overlap fully. One possible solution would be to use a smaller pixel size, which would position the pixels closer together and increase the overlap of the high frequency components in the far-field. This may also benefit the 

In conclusion, a novel photoconductive emitter geometry for broadband THz polarization rotation, consisting of separate interdigitated pixels for emission of horizontal and vertical polarization states, has been proposed and tested. The generated orientation angle of the THz pulse was shown to be controllable by varying the relative bias voltages applied to the horizontally and vertically emitting contacts, and remained close to the target values over a $360^{\circ}$ rotation of the polarization state. Calculations of the emitted radiation pattern showed that in the far-field beams from the individual pixels have overlapped, and demonstrated that the \textcolor{black}{calculated} gaussian beam quality of the emitter is higher at lower frequency. \textcolor{black}{Future experiments will investigate the radiation pattern experimentally.} The wide bandwidth ($0.3-5.0$\,THz), scalable design, and simple polarization rotation method (varying the applied bias voltage to each contact) make this emitter geometry an attractive concept for applications in spectroscopy and imaging systems, in which the THz polarization state may be rotated arbitrarily without resorting to any physically moving parts. Since the polarization angle of the THz radiation generated by this device depends only upon the applied bias voltages, the THz polarization can be modulated at speeds much faster than systems relying on mechanically rotating components, such as polarizers, which are limited to low frequencies (e.g.\ 15\,Hz).\cite{Aschaffenburg2012} The general concept of pixel-based photoconductive devices also has further applications in THz detection, for spectroscopy, imaging and THz beam profiling.

The authors would like to thank Michele Failla, Hugh Thomas, Lucas Bartol-Bibb and Mark Crouch for advice and technical assistance. The authors acknowledge funding from the EPSRC (UK). Data related to this publication is available from the University of Warwick data archive at http://wrap.warwick.ac.uk/113369.
\\


\end{document}